\documentclass[aps,pra,preprint,groupedaddress,showpacs]{revtex4}
\begin{document}

\title{The Effect of Noise on the Dirac Phase of Electron in The Presence of Screw Dislocation}

\author{Reza Torabi}
\affiliation{Physics Department, Tafresh University of Technology,
Tafresh, Iran} \email{rezatorabi@aut.ac.ir}

\begin{abstract}
The effect of noise on the Dirac phase of electron in the presence
of screw dislocation is studied. An uncorrelated noise, which
coincides with the nature of thermal fluctuations, is adopted.
Results indicate that the Dirac phase is robust against existing
noise in the system.
\end{abstract}

\pacs{03.65.Vf, 61.72.Lk, 05.40.-a, 02.40.Ky}

\maketitle

\section{Introduction}

Some of solids have a crystiline structure which is not an ideal
structure and has defects. These defects have strong effect on
physical properties of the medium
\cite{Mehrafarin,Mesaros,Teichler,Kirova,Giannattasioa,Osipov,Vozmediano,Brandenberger,Furtado,Furtado2,Yazyev,Bausch,Bausch2}.
Therefore, study them is important in solid state physics.

Elastic stresses in solids due to the defects mathematically
corresponds to a non-Euclidean space. This is the Katanaev-Volovich
approach to the theory of defects in solids which is analogous to
three dimensional gravity \cite{Katanaev,Kleinert}. Dealing with a
non-Euclidean metric is easier than using complicated boundary
conditions. In the continuum limit, which is valid at distance much
larger than the lattice spacing, the solid is described by a
Riemann-Cartan manifold with curvature and torsion. Topological
defects such as disclination and dislocation are described by
curvature and torsion, respectively.

In 1931, Dirac \cite{Dirac} showed that when a particle transports
in an external electromagnetic field, its wave function acquires a
phase term in addition to usual dynamic phase factor. The change of
the phase under the transfer along a close counter is proportional
to the field flux through the counter. This additional phase is
known as Dirac phase and is a non-integrable phase factor
\cite{Bliokh}. Non-integrable phase factor appears in many different
area of physics \cite{Shapere}.

Furtado and coworkers \cite{Furtado3} investigated the Dirac phase
of electron in the presence of screw dislocation using the
Katanaev-Volovich approach and showed that it is analogous to
Aharonov-Bohm effect \cite{Aharonov}. In this case the Burger vector
of dislocation plays the role of electromagnetic flux.

In this paper, we study the effect of noise, which usually presents
in physical systems, on the dirac phase of electrons in media with
screw dislocation. Theoretical results indicate that the dirac phase
factor of screw dislocation is robust against fluctuations.

The paper is organized as follow. In section II, we review the Dirac
phase of electrons in the presence of screw dislocation. In section
III, The effect of noise on the mentioned Dirac phase factor is
studied. The selected model of noise coincides with the nature of
thermal fluctuations. At the end, the conclusion is presented.

\section{Dirac phase in the presence of screw dislocation}

In a screw dislocation the Burger vector is parallel to the
dislocation line. This type of defect, which corresponds to a
singular torsion \cite{Bilby} along the defect line, is described by
the following metric \cite{Tod,Furtado3}
\begin{equation}
ds^{2} =g_{ij}dx^{i}dx^{j}= \left( dz + \beta d \phi \right)^{2} + d
\rho^{2} + \rho^{2} d \phi^{2},
\end{equation}
where $\beta$ is a parameter related to the Burger vector $\vec{b}$
by $\beta = \frac{b}{2 \pi}$. This metric carries torsion but no
curvature.

The schr\"{o}dinger equation for an electron in the presence of
defect is given by
\begin{equation}
-\frac{\hbar^2}{2m} \nabla^2 \psi(\rho, \varphi, z,t)=i\hbar
\frac{\partial}{\partial t} \psi(\rho, \varphi, z,t),
\end{equation}
where the Laplace-Beltrami operator is given by
$\nabla^2=\frac{1}{\sqrt{g}}\partial_i(g^{ij}\sqrt{g}\partial_j)$.
$g$ is the determinant of the metric tensor $g_{ij}$. The metric
tensor $g_{ij}$ and its inverse $g^{ij}=(g_{ij})^{-1}$ are as follow
\begin{equation}
g_{ij}=\left( \begin{array}{ccc} 1 & 0 & 0 \\
0 & \beta^2+\rho^2 & \beta \\
0 & \beta & 1 \end{array} \right),\;\;
g^{ij}=\left({{\begin{array}{ccc} 1 & 0 & 0 \\
0 & \frac{1}{\rho^2} &
 -\frac{\beta}{\rho^2} \\
 0 & -\frac{\beta}{\rho^2} & \frac{\beta^2+\rho^2}{\rho^2}\\
\end{array} }} \right).
\end{equation}

Using the metric tensor (3) for screw dislocation, the
schr\"{o}dinger equation (2) takes the form
\begin{equation}
-\frac{\hbar^2}{2m}
\bigg{\{}\partial_z^2+\frac{1}{\rho}\partial_\rho(\rho\partial_\rho)+\frac{1}{\rho^2}(\partial_\varphi-\beta\partial_z)^2\bigg{\}}
\psi(\rho, \varphi, z,t)=i\hbar \frac{\partial}{\partial t}
\psi(\rho, \varphi, z,t),
\end{equation}
whose solution can be written as
\[
\psi(\rho,\varphi,z,t)=e^{-\frac{iEt}{\hbar}}e^{ikz}\psi(\rho,\varphi).
\]
In this way the schr\"{o}dinger equation (4) is given by
\begin{equation}
-\frac{\hbar^2}{2m}
\bigg{\{}-k^2+\frac{1}{\rho}\partial_\rho(\rho\partial_\rho)+\frac{1}{\rho^2}(\partial_\varphi-i\beta
k)^2\bigg{\}} \psi(\rho, \varphi)= E \psi(\rho, \varphi),
\end{equation}

Equation (5) implies that
$\partial_\varphi\rightarrow\partial_\varphi\ -ik\beta$ with respect
to the defect free case ($\beta=0$) in which the Laplacian operator
is given in a flat space. In the other words, the angular momentum
changes according to $l \rightarrow\ l -k\beta$. The angular
momentum of the electron is modified by the presence of the defect
which is due to the torque exerted by the strain field of the
dislocation. Thus, an electron in the presence of screw dislocation
behaves like an electron in the presence of a magnetic flux. The
corresponding vector gauge potential is
\begin{equation}
\mathbf{A}=\frac{k\beta}{\rho} \hat{\mathbf{e}}_\varphi.
\end{equation}
According to this correspondence, the Dirac phase factor method
\cite{Dirac,Sakurai} can be used. Therefore, the solution of the
schr\"{o}dinger equation (5) can be written as
\begin{equation}
\psi(\rho,\varphi)=\exp\bigg{\{}i\int \mathbf{A}\cdot
d\mathbf{r}\bigg{\}}\psi_0(\rho,\varphi),
\end{equation}
where $\psi_0(\rho,\varphi)$ is the solution of the defect free
case. Substituting  (6) in (7) leads to
\[
\psi(\rho,\varphi)=e^{i\int_{\varphi_0}^{\varphi}k\beta
d\varphi}\psi_0(\rho,\varphi).
\]
This means that $\psi(\rho,\varphi)$ differs from
$\psi_0(\rho,\varphi)$ just in a factor $e^{i\gamma}$, where
\begin{equation}
\gamma=\int_{\varphi_0}^{\varphi}k\beta d\varphi,
\end{equation}
and is known as Dirac phase factor. This phase depends on electron
trajectory.

\section{The effect of noise on the Dirac phase}

The presence of noise is an unavoidable subject in physical systems.
The noise for example, can be due to ubiquitous thermal fluctuations
in the system. In this section we study the effect of noise on the
Dirac phase factor. The noise does not affect the the magnitude of
wave vector, $k$. The only effect of noise is to change Burger
vector according to
\begin{equation}
\mathbf{b}(t)=\mathbf{b}_0+\mathbf{N}(t),
\end{equation}
where the index $"0"$ indicates the absence of noise and
$\mathbf{N}(t)$ is the noise term. The noise is a random process
with zero average and small amplitude compared to $\mathbf{b}_0$.
So, the effect of noise on the metric is negligible due to the
smallness of the noise amplitude. Naturally, the metric can be
considered like (1) just by having beta as a function of time,
$\beta(t)=\frac{b(t)}{2\pi}$. Consequently, by following the same
procedure, $\gamma$ leads to the result (8) while beta is now time
depenndent and the electron trajectory is a fluctuating one. It
fluctuates about the noiseless trajectory of the electron (assumed
cyclic with period $T$). Using (9) the resulting change in $\gamma$
during time $T$ will be
\[
\Delta \gamma=\frac{k}{T}\int_0^T (b-b_0)dt.
\]
In deriving above we used $d\varphi=\frac{2\pi}{T}dt$. $\mathbf{b}$
does not return to its original direction because of the random
noise and a non-cyclic contribution also appears. According to the
definition of non-integrable phase for non-cyclic evolution
\cite{Samuel}, this term can be removed and the above result still
holds \cite{Chiara}.

Expanding $b=|\mathbf{b}_0+\mathbf{N}|$ in terms of $\mathbf{b}_0$
to the first order in the noise, yields
\begin{equation}
\Delta \gamma=\frac{k}{T}\int_0^T
\frac{\mathbf{b}_0\cdot\mathbf{N}}{\mathbf{b}_0}dt.
\end{equation}
The noise has zero average, thus $<\Delta\gamma>=0$. It means that
the average value of $\gamma$ coincides with its noiseless value. In
order to compute the probability distribution of $\Delta\gamma$, a
definite model for the noise is needed. We want to compute the
second moment of $\Delta\gamma$ which is related to the width of the
probability distribution function.

A suitable model for noise with respect to the physical nature of
thermal fluctuations is the uncorrelated noise defined by
\[
<\mathbf{N}_i(t)>=0,\\ \\
<\mathbf{N}_i(t)\mathbf{N}_j(\acute{t})>=2D\delta(t-\acute{t})\delta_{ij},
\]
where angular bracket denotes ensemble average. Using this
definition for noise, the second moment of $\Delta\gamma$ which is
defined in (10) leads to
\[
<\Delta\gamma^2(T)>=\frac{k^2}{T}.
\]
It increases with second power of wave vector and means that the
probability distribution function for the change of $\gamma$ phase,
which reflects the effect of fluctuations, decreases by choosing low
energy electrons. Also it slows down in time by the factor
$\frac{1}{T}$ so $\gamma$ coincides with its noiseless value in the
adiabatic limit $T\rightarrow \infty$. As a result, in spite of
dynamic phase, the Dirac phase is robust against existing noise in
the system .

\section{Conclusion}

The effect of noise, which is presented in physical systems, on the
Dirac phase of electrons in media with screw dislocation is studied.
The selected noise is an uncorrelated random noise which is
coincided with the nature of thermal fluctuations. The second moment
of the change in the Dirac phase, which is related to the width of
the probability distribution function, slows down in time by the
factor $\frac{1}{T}$. It indicates that dirac phase is robust
against fluctuations.

\end{document}